\documentclass[aps,twocolumn,groupedaddress,floatfix,superscriptaddress]{revtex4}

\begin{document}
\bibliographystyle{apsrev}

\title{Particle-hole symmetry and the Pfaffian state}

\author{Michael Levin}
\author{Bertrand I. Halperin}
\author{Bernd Rosenow}
\affiliation{Department of Physics, Harvard University, Cambridge, Massachusetts 02138}

\date{\today}

\begin{abstract}
We consider the properties of the Moore-Read Pfaffian state under particle-hole conjugation.
We show that the particle-hole conjugate of the Pfaffian state - or ``anti-Pfaffian" state -
is in a different universality class from the Pfaffian state, with different topological order.
The two states can be distinguished by both their bulk and edge physics though the difference is 
most dramatic at the edge: the edge of the anti-Pfaffian state has a composite structure that 
leads to a different thermal Hall conductance and different tunneling exponents than the 
Pfaffian state. At the same time, the two states are exactly degenerate in energy for a $\nu 
= 5/2$ quantum Hall system in the idealized limit of zero Landau level mixing. Thus, both 
are good candidates for the observed $\sigma_{xy} = \frac{5}{2}(e^2/h)$ quantum Hall 
plateau. 
\end{abstract}

\maketitle

\textsl{Introduction}:
The Moore-Read Pfaffian ($Pf$) state \cite{MR9162} is believed to be a strong candidate for the observed
quantum Hall plateau with $\sigma_{xy} = \frac{5}{2}(e^2/h)$. \cite{XPV0409} This possibility is 
particularly exciting since the quasiparticle excitations in this state carry non-abelian statistics. 
Much work has been devoted to understanding the basic physical properties of the $Pf$ state. 
However, one aspect of the $Pf$ state has not been addressed - namely, its behavior under particle-hole (PH) 
conjugation of electrons in the spin-aligned partially occupied second Landau level. This issue is 
important because, to a good approximation, the Hamiltonian of $\nu = 5/2$ FQH system is symmetric 
under this PH conjugation. Indeed, this symmetry is exact if one considers a model with no Landau level 
mixing, complete spin alignment in the second Landau level, only two body interactions, and precisely 
one-half electron per flux quantum.

On the other hand, the $Pf$ state is \emph{not} particle-hole 
symmetric: it is the exact ground state of a three body Hamiltonian that explicitly breaks 
particle hole symmetry. \cite{GWW9105} Therefore, if the ground state of the idealized $\nu 
= 5/2$ model is well represented by the Pfaffian (as numerical studies suggest 
\cite{MN0309,RH0085}) then we are led to an interesting conclusion: this model must 
spontaneously break PH symmetry. In addition to the Pfaffian state, there must be 
another exactly degenerate ground state: the particle-hole conjugate of the Pfaffian, or 
``anti-Pfaffian" ($APf$) state.  

We shall argue below that the $Pf$ and $APf$ states represent topologically distinct phases: one 
cannot go from one to another (at $T=0$) by varying a parameter (such as the sign of three-body 
interaction) without encountering a phase transition.
The easiest way to demonstrate this is to consider the ``edges" of the two 
states; i.e., the boundary between these states and the ``vacuum" $\nu=2$ 
state.
We shall see that the edge structure is 
necessarily different for the two states. (The underlying PH symmetry implies, however, that the 
boundary between a $Pf$-like $\nu=5/2$ state and $\nu=3$ is equivalent 
to the edge between an $APf$-like $\nu=5/2$ state and $\nu=2$, and vice versa). 

Of course, exact PH symmetry is only a property of the idealized model at Landau level filling  
factor $f=5/2$. In the real system, there will be terms which break the symmetry, such as three-body 
interactions which arise in second-order
perturbation theory from Landau-level mixing. The sign of 
these symmetry breaking terms is not currently known, but presumably they favor one of the two 
states, $Pf$ or $APf$, at $f=5/2$. 
If the symmetry breaking terms are large enough, they may stabilize one 
of these states throughout the entire range of filling fractions (roughly $2.49 < f < 2.51$ ) where the 
$\nu=5/2$ Hall plateau has been observed.  

On the other hand if the symmetry-breaking terms are small, there may be a phase 
transition between $Pf$ and $APf$ somewhere in the range $2.49 < f < 2.51$. Indeed, away from $f=5/2$, 
there will be a net concentration of localized quasiparticles or quasiholes, which presumably have 
different energies in the $Pf$ and $APf$ states. This energy difference can overcome the splitting due to
symmetry-breaking for sufficiently large deviations from half-filling. A natural guess for the
sign of this effect is that quasiholes have lower energy in the $Pf$ state, 
(since the $Pf$ edge has a simpler edge structure, as we shall see below) while quasiparticles have lower 
energy in the $APf$ state. Such an asymmetry favors $Pf$ for $f<2.5$ and $APf$ for $f>2.5$.

If there is a phase transition at a filling fraction $f_c$ in range $2.49 < f < 2.51$, then taking 
disorder into account, one might expect there to be a region close to $f_c$ where both phases exist, 
with a network of domain walls separating them. The energy gap for charged excitations should be 
reduced, or perhaps vanish, near these walls, which should lead to a peak in the resistivity for $f 
\approx f_c$. To the best of our knowledge, such a resistivity peak in the middle of the $\nu=5/2$ 
Hall plateau has not been seen in existing experiments. This suggests that one of the two states, $Pf$ 
or $APf$, has been stabilized throughout the region. 


In the remainder of this letter, we address the most important differences between the $Pf$ and $APf$ 
states. We analyze the $APf$ edge and show that it has different thermal Hall 
conductance $K_H$ and different tunneling exponents from the $Pf$ edge. In addition, while 
the quasiparticles in the $Pf$ and $APf$ states have similar non-abelian statistics, we show that they 
differ in their detailed structure - another demonstration that the two states are 
topologically distinct.

\textsl{Thermal Hall conductance}: 
In this section we show that the $APf$ edge has a different thermal Hall conductance than the $Pf$ 
edge - implying that the two states have different topological orders and are in different 
universality classes. 

The definition of the thermal Hall conductance is similar to the electric Hall 
conductance. Just as an electric current flows when the top and bottom edges of a quantum 
Hall bar are at different voltages, there will generally be a net transport of energy along 
the Hall bar when the top and bottom edges are at different temperatures. The thermal Hall 
conductance is defined by $K_H = \frac{\partial J_{Q}}{\partial T}$, where $J_Q$ is the thermal energy 
current carried by the edge modes.

The thermal Hall conductance is an important quantity because, like the electric Hall 
conductance, it is universal. It is insensitive to interactions and disorder at the edge and 
depends only on the topological order of the bulk quantum Hall states. \cite{KF9732,K0538} 
Indeed, conservation of energy implies that if there is an energy gap in
the bulk, $K_H$ cannot vary from one region of a Hall bar to another (unless there is a phase boundary
with low energy excitations which crosses from one edge to the other). 


In addition, it is simple to compute once one knows the edge theory. One can show that each 
chiral boson edge mode gives a contribution $\pm \frac{\pi^2 k_B^2}{3h}T$ to the thermal 
Hall conductance, with the sign determined by the direction of propagation. A Majorana fermion mode if 
present, contributes half as much as a boson mode. 

To proceed further we now construct an edge theory for the $APf$ edge. As we are only 
interested in universal quantities at the moment, we may use the simplest possible model for the 
edge. We consider only electrons in the second Landau level so we discuss a model for the edge between 
$\nu=0$ and $\nu=1/2$. In the simplest model we assume there is a narrow region of filling fraction 
$\nu = 1$ separating the $\nu = 1/2$ $APf$ bulk and the $\nu = 0$ vacuum. In 
this scenario, the edge theory can be decomposed into two pieces - an ($APf$, $\nu = 
1$) edge and a $(\nu = 1, \nu = 0)$ edge. The ($APf$, $\nu = 1$) edge is equivalent via 
particle-hole symmetry to a ($Pf$, $\nu = 0$) edge. It therefore consists of a backward 
propagating Majorana fermion $\psi$ and a backward moving charged chiral boson $\phi_1$. \cite{W9355} 
(We use the convention that the ``forward" direction is clockwise for $B \parallel 
\hat{z}$). The $(\nu = 1, \nu = 0)$ edge is a simple IQH edge and consists of a single 
forward propagating chiral boson $\phi_2$. The appropriate Lagrangian is thus $\mathcal{L} = 
\mathcal{L}_1 + \mathcal{L}_2 + \mathcal{L}_{12}$ where
\begin{eqnarray}
\mathcal{L}_1 &=& i\psi(\partial_t \psi + v_n \partial_x \psi)
+\frac{2}{4\pi}\partial_x \phi_1 (-\partial_t \phi_1 - v_1 \partial_x 
\phi_1)
\nonumber \\
\mathcal{L}_2 &=& \frac{1}{4\pi}\partial_x \phi_2
(\partial_t \phi_2 - v_2 \partial_x \phi_2)
\nonumber \\
\mathcal{L}_{12} &=& -\frac{2}{4\pi} v_{12} \partial_x \phi_1 \partial_x 
\phi_2
\label{cleanedge}
\end{eqnarray}
and $\mathcal{L}_{12}$ encodes the repulsive density-density interaction between the two 
branches. 

To compute the thermal Hall conductance, we only need to count the number of forward and back 
propagating modes. In the above edge theory (\ref{cleanedge}) there is one backward 
propagating Majorana fermion, one backward propagating chiral boson, and one forward 
propagating chiral boson. The thermal Hall conductance is therefore $\tilde{K}_H = -1/2 - 1 
+ 1 = -1/2$ (in units of $\frac{\pi^2 k_B^2}{3h}T$). On the other hand, the thermal Hall 
conductance of the Pfaffian edge is $K_H = 1/2 + 1 = 3/2$ (as it consists of a forward 
propagating Majorana fermion and a forward propagating chiral boson). Since $\tilde{K}_H \neq K_H$, we 
conclude that the two states must be in different universality classes. 

An important feature of of the $APf$ edge is that it is necessarily a
\emph{composite} edge. That is, it always has at least two modes moving in opposite directions, in 
addition to the required Majorana mode. This result cannot be affected by any edge-reconstruction or 
disorder. Indeed, suppose there were only one chiral boson mode and one 
Majorana mode with directionalities $\eta_1, \eta_n = \pm 1$. In order to get the correct 
thermal Hall conductance we must have $\eta_1 + \eta_n/2 = -1/2$; at the same time the sign 
of the electric Hall conductance implies $\eta_1$ is positive. These two constraints are 
incompatible. Satisfying them requires introducing an additional counter-propagating mode. 

\textsl{Random edge}: While the clean edge theory (\ref{cleanedge}) captures the universal 
properties and basic mode structure of the minimal $APf$ edge, it is missing a 
crucial dynamical feature: it contains no mechanism for equilibration between the forward 
propagating $\psi$ and $\phi_1$ modes and the backward propagating $\phi_2$ modes. In this 
section we consider the effect of such physics. Our analysis closely follows previous work on 
the disordered $\nu = 2/3$ edge. \cite{KFP9429}

To describe such equilibration physics, we need to include a tunneling process in which an 
electron tunnels from the $\psi, \phi_1$ edge to the $\phi_2$ edge. In the above notation, 
the operator $\psi e^{2i\phi_1}$ creates an electron on the $\psi,\phi_1$ edge while the 
operator $e^{-i\phi_2}$ creates an electron on the $\phi_2$ edge. Thus, the simplest and 
most relevant tunneling term is $\psi e^{2i\phi_1 +i\phi_2}$ - which moves a single electron 
from the $\phi_2$ edge to the $\psi,\phi_1$ edge. 

In a realistic edge we expect that such a tunneling term will come in with a coefficient $\lambda(x)$
which contains a constant part $\lambda_0$ and a spatially varying part $\delta \lambda$, due to 
disorder. In the general case, the two charge modes correspond to electrons with different momenta, and 
the term $\lambda_0$ will be unimportant, if it is sufficiently weak. The random part can be important, 
however, if it has Fourier components that match the momentum difference between the two modes. 
Thus, we are led to consider only the tunneling due to disorder - which we model by
\begin{equation}
\mathcal{L}_{tun} = \psi e^{2i\phi_i + i\phi_2}\xi(x) + h.c 
\end{equation}
where $\xi$ is a complex Gaussian random variable satisfying 
$\<{\xi(x) \xi^*(x')}\> = W \delta (x-x')$.

The effect of weak disorder $W$ can be analyzed by computing the scaling dimension of the 
above tunneling operator. The result is
\begin{equation}
\Delta = \frac{1}{2} + \frac{3/2 - 2x}{\sqrt{1-2 x^2}}
\end{equation}
where $x = v_{12}/(v_1+v_2)$. To determine whether the tunneling term is relevant or not, we 
need to analyze its behavior under an RG flow. The appropriate RG flow equation for a 
spatially random perturbation in dimension $1$ is $dW/dl = (3-2\Delta)W$. \cite{GS8825} We 
conclude that disorder is irrelevant for $\Delta > 3/2$ and relevant for $\Delta < 3/2$. The two cases 
correspond to two different phases for the $APf$ edge. In the first 
phase, the edge physics is captured by the clean edge theory (\ref{cleanedge}). In this 
phase the edge modes do not equilibrate at $T=0$. Much of the physics - including the values of 
tunneling exponents - is non-universal. In the second phase, disorder mediated 
equilibration plays an important role. In the next section, we analyze this disorder dominated phase.

\textsl{Disorder dominated phase with emergent $SO(3)$ symmetry}: It is convenient to make 
a change of basis from the $\phi_1,\phi_2$ fields to a neutral field $\phi_\si$ and a 
charged field $\phi_\rho$ defined by
\begin{equation}
\phi_\si = 2\phi_1 + \phi_2, \ \ \ \phi_\rho = \phi_1 + \phi_2
\end{equation}
Rewriting the Lagrangian in terms of the new fields, we find 
$\mathcal{L} = \mathcal{L}_{\rho} + \mathcal{L}_{\si} + 
\mathcal{L}_{\rho\si}$ where 
\begin{eqnarray}
\mathcal{L}_\rho &=& \frac{2}{4\pi}\partial_x \phi_\rho(\partial_t
\phi_\rho- v_\rho \partial_x \phi_\rho) \nonumber \\
\mathcal{L}_\si &=& \frac{1}{4\pi}\partial_x \phi_\si
(-\partial_t \phi_\si - v_\si \partial_x \phi_\si) 
+ i\psi(\partial_t \psi + v_n \partial_x \psi)
\nonumber \\
\mathcal{L}_{\rho\si} &=& -\frac{2}{4\pi}v_{\rho\si} \partial_x
\phi_\rho \partial_x \phi_\si 
\label{bosedge}
\end{eqnarray}
The tunneling can also be rewritten as
\begin{eqnarray}
\mathcal{L}_{tun} &=& \psi e^{i\phi_\si}\xi(x) + h.c
\end{eqnarray}

It is useful to think of the chiral boson $\phi_\si$ as a chiral fermion $\psi_\si = 
e^{i\phi_\si}$. This fermion can in turn be thought of as two Majorana fermions. Thus 
altogether, the edge contains three backward propagating neutral Majorana fermions along 
with a forward propagating charged boson $\phi_{\rho}$. This triplet of Majorana fermions is 
defined by $\psi_1 = \psi$, $\psi_2 = e^{i\phi_\si}+e^{-i\phi_\si}$, $\psi_3 = 
-i(e^{i\phi_\si} -e^{-i\phi_\si})$.

The physics of the edge is particularly simple in the Majorana fermion description. Consider 
the tunneling term $\mathcal{L}_{tun}$. It can be written
\begin{equation}
\mathcal{L}_{tun} = \psi^T (\vec{\xi}(x) \cdot \vec{L})\psi
\end{equation}
where ${\vec{\xi}} = Re(\xi)\hat{y}/2 + Im(\xi)\hat{z}/2$ and $L_x,L_y,L_z$ are the three 
generators of $SO(3)$. The effect of this coupling is to mix the three modes so that, at 
length scales larger than the mean free path, they propagate with the same average velocity 
$\bar{v} = (v_n + 2 v_\si)/3$. The other effect 
of the tunneling term is to suppress the coupling $\mathcal{L}_{\rho\si}$ between the three Majorana 
modes and the charged boson $\phi_\si$. Indeed, $\mathcal{L}_{\rho\si}$ breaks the $SO(3)$ 
symmetry between the modes and thus ``averages to zero" in the presence of the random mixing 
term $\mathcal{L}_{tun}$. 
The end result is that the three Majorana modes decouple from the charged boson 
$\phi_\rho$ and propagate at the same velocity, $-\bar{v}$. The disorder dominated edge thus 
contains an emergent $SO(3)$ symmetry.

To understand this physics more quantitatively, we rewrite the Lagrangian in terms of the 
three Majorana fields. The result is 
$\mathcal{L} = \mathcal{L}_\rho+\mathcal{L}_0 +\mathcal{L}_{anis}$ where
\begin{equation}
\mathcal{L}_0 = 
i \psi^T (\partial_t \psi + \bar{v}\partial_x \psi -i 
(\vec{\xi}(x) \cdot \vec{L})\psi)
\end{equation}
describes an $SO(3)$ symmetric triplet of Majorana fermions with random quadratic coupling 
and $\mathcal{L}_{anis}$ contains the anisotropic terms
\begin{equation}
\mathcal{L}_{anis} =  i \psi^T (\delta v) \partial_x \psi -  
\partial_x \phi_{\rho}\psi^T (\vec{v}_I \cdot \vec{L}) \psi
\label{anis}
\end{equation}
Here, $\delta v$ is the velocity-anisotropy matrix $\delta v = 
\text{diag}(v_n-\bar{v},v_\si-\bar{v},v_\si-\bar{v})$ and $\vec{v}_I = v_{\rho \si} \hat{x}/8\pi$.

In the following, we argue that, because of the random coupling $\vec{\xi}(x)$, the anisotropic 
couplings $\mathcal{L}_{anis}$ are irrelevant. Thus, at least for small anisotropy $v_n 
\approx v_{\si}$ and small coupling $v_{\rho\si}$, the disordered $APf$ edge will flow to the 
$SO(3)$ symmetric fixed line described by $\mathcal{L} = 
\mathcal{L}_\rho+\mathcal{L}_0$. To see this, it is useful to make the change of variables 
$\psi'(x) = R(x) \psi(x)$ where
\begin{equation}
R(x) = P \exp(-\frac{i}{\bar{v}}\int_{-\infty}^x dy \ \vec{\xi}(y) \cdot 
\vec{L})
\end{equation}
and $P$ denotes the path ordering operator. In terms of the rotated fields $\psi'$, $\mathcal{L}_0$ is 
simply
\begin{equation}
\mathcal{L}_0 = i \psi'^T (\partial_t \psi' + \bar{v} \partial_x \psi')
\end{equation}
Thus, the fixed line Lagrangian contains three Majorana fermion modes propagating at the 
average velocity $-\bar{v}$, and a completely decoupled charged boson $\phi_\rho$.

Now consider the effect of small anisotropy $\mathcal{L}_{anis}$. Writing 
$\mathcal{L}_{anis}$ in terms of the $\psi'$ fields gives
\begin{eqnarray}
\mathcal{L}_{anis} &=&  i \psi'^T (\delta v'(x)) \partial_x 
\psi' - \partial_x \phi_{\rho}\psi'^T (\vec{v}'_I(x) \cdot \vec{L}) 
\psi' \nonumber \\
&-&\psi'^T (\frac{\del v}{\bar{v}} \vec{\xi}(x) \cdot \vec{L} )\psi'
\label{anis2}
\end{eqnarray}
where the the effective couplings $\delta v'(x) = R(x) \delta v R^{-1}(x)$, $\vec{v}'_I(x) = R(x) 
\vec{v}_I$ are now spatially dependent. 

It is not hard to see that these terms are irrelevant. Consider, for example, the first 
term, $\psi'^T\partial_x \psi'$. Its scaling dimension is $\Delta = 2$. On the 
other hand, its spatial coefficient is not uniform: 
$\del v'(x)$ is spatially dependent and random due to the random nature of the rotation 
$R(x)$. Thus the appropriate RG flow equation for this term is the flow equation for the 
\emph{mean square average} $W_{\del v'} =\<\del v'(x)^2\>$. Since this equation is 
$dW_{\del v'}/d l = (3 - 2\Delta)W_{\del v'}$ \cite{GS8825}, the first term is irrelevant. In the same 
way, one can show that the second term in (\ref{anis2}) is irrelevant. As for the third term, it is 
unimportant since it simply renormalizes the disorder $\vec{\xi}(x)$. 

\textsl{Tunneling exponents for a point contact}:
In this section we compute tunneling exponents for the $APf$ edge in the disorder 
dominated phase, and compare with the corresponding exponents for the $Pf$ edge. 
To begin, consider the most general quasiparticle creation operator for the anti-Pfaffian 
edge (\ref{cleanedge}). From the decomposition of (\ref{cleanedge}) into the $Pf$-like 
edge $\mathcal{L}_1$ and the IQH edge $\mathcal{L}_2$, the most general creation operator is of the 
form $\mathcal{O}_{\chi,n_1,n_2} = \chi e^{i n_1 \phi_1 + i n_2 \phi_2}$ where $n_2$ is an integer, 
$\chi = I,\si,\psi$ ranges over the three fields in the chiral Ising model, and $n_1$ is either an 
integer or a half-integer depending on whether $\chi = 1,\psi$ or $\chi = \si$. This operator creates 
an edge excitation with charge $q = (n_1/2 - n_2) e$.

The relevant quantity for analyzing tunneling across a point contact is the exponent $g$ 
defined by $\<\mathcal{O}(x,t=0)\mathcal{O}(x,t)\> \sim t^{-g}$. \cite{W9505} This quantity 
can be computed easily along the $SO(3)$ symmetric fixed line. The reason is that both 
operators are measured at the same point $x$ and thus the random rotation $R(x)$ does not 
enter into the problem. We can thus ignore disorder and compute the correlation function in 
the original bosonic representation of the edge Lagrangian (\ref{bosedge}), setting 
$v_{\rho\si} = 0$, $v_{\si} = v_{n}$, $\mathcal{L}_{tun} = 0$. We find $g(\chi,n_1,n_2) = 
g_\chi + (n_1 - n_2)^2 + (2 n_2 - n_1)^2/2$, where $g_{I} = 0, g_{\si} = 1/8, g_{\psi} = 
1$.

We now use this result to compute the temperature dependence of the tunneling current
$\del I$ across a weak constriction. The tunneling current is controlled by 
processes where quasi-particles tunnel from one edge 
of the Hall bar to the other. The dominant contribution comes from the (charged) 
quasi-particle creation operator with the smallest exponent $g$. For our case, the 
smallest exponent is $g_q = 1/2$, obtained by $\si e^{i\phi_1/2}$ (as well as other 
operators). The temperature dependence of the tunneling current (for voltage $V \ll k_B T/e$) is 
therefore $\del I \sim T^{2 g_q -2} = T^{-1}$ \cite{W9505}. Note that this is different from the 
result for the $Pf$ edge. In that case, $g_q = 1/4$ so the tunneling current behaves as 
$\del I \sim T^{-3/2}$. \cite{W9355}

One could also consider the limit of a strong constriction where the relevant tunneling 
current is between two separate $APf$ droplets. However this geometry is not 
as useful for distinguishing the two states. Indeed, in this limit, the tunneling 
current is controlled by processes where \emph{electrons} tunnel from one anti-Pfaffian 
droplet to the other. One can check that in both the $Pf$ and $APf$ states the 
smallest exponent for an electron creation operator is $g_e = 3$. Thus, in both cases the
temperature dependence of the tunneling current is $I_{tun} \sim T^{2 g_e-2} = T^4$. 
\cite{W9355} 

Recently, Miller et al \cite{MRZ2007} observed a zero bias resistance peak for current flowing 
through an $f \approx 5/2$ constriction, separating two $\nu = 3$ bulk regions. Given the exponents
found above (in particular, the fact that $g_{q,Pf} < g_{q,APf}$), and using a simple model
for the constriction \cite{L0618}, this peak appears consistent with a $Pf$-like 
$5/2$ state but not an $APf$-like $5/2$ state. In the latter case, ``forward 
scattering" of quasiparticles between the two boundaries with $\nu=3$ should be more important at 
low voltages than ``back-scattering" between the vacuum edges, leading to a zero bias \emph{minimum}.

\textsl{Quasiparticle statistics}:
In this section we investigate the properties of the quasiparticle excitations in the 
$APf$ state. To this end, note that each excitation $\tilde{a}$ of the 
$APf$ state is particle-hole conjugate to an excitation $a$ of the $Pf$ state. 
From the properties of particle-hole symmetry, $\tilde{a}$ carries the opposite charge from 
its particle-hole conjugate: $q_{\tilde{a}} = -q_{a}$. Also, it carries the opposite 
braiding statistics: $\theta_{\tilde a \tilde b} = - \theta_{a b}$ where
$\theta_{\tilde{a}\tilde{b}}$ ($\theta_{a b}$) is the phase associated with braiding 
$\tilde{a}$ ($a$) around $\tilde{b}$ ($b$). These relations completely determine the 
excitation spectrum for the $APf$ state. Its basic structure is quite similar to 
the $Pf$ state - it contains a charge $e/4$ quasihole $\tilde{\si}$ with non-abelian 
statistics, a neutral fermion $\tilde{\psi}$, etc.

However, we would like to emphasize that the excitation spectrum is \emph{physically 
distinguishable} from the $Pf$ state. For example, one can check that the phase 
associated with exchanging two charge $e/4$ quasiholes is different for the two states. In the $Pf$ 
state, the phase is $1$ if the two quasiholes are in the $I$ fusion channel and 
$i$ if they are in the $\psi$ fusion channel \cite{NW9629}; in the $APf$ state, the 
phase is again $1$ in the $I$ channel, but it is $-i$ in the $\psi$ channel (for an 
appropriate orientation convention). This difference in quasiparticle statistics is 
measurable in principle, and provides another distinction between the topological orders 
in the two states. However it is a subtle effect that would not be detectable in the 
recently proposed interferometry experiments, \cite{SH0602, BKS0603}.

\textsl{Implications for numerics}: Suppose the ground state of the idealized PH-symmetric model
for $\nu = 5/2$ is $Pf$-like (or $APf$-like). Then, a finite size calculation on a torus with an even number 
of electrons should yield $2 \cdot 6 = 12$ nearly degenerate ground states - $6$ corresponding 
roughly to ``even" combinations of $Pf$-like and $APf$-like states, and $6$ corresponding to ``odd" 
combinations. Numerical calculations are consistent with this picture, but for the largest sizes studied 
to date, the energy gap separating the two sets of $6$ states is not small and it has not been 
possible to check that the gap diminishes with increasing system size. \cite{EHR2007}
Note that this effect is not present on a finite sphere, since in that geometry the $Pf$ and $APf$ states 
actually occur at different electron numbers. \cite{MN0309}

\textsl{Acknowledgments}:
The authors are grateful for helpful discussions with A. Stern, S. H. Simon, C. M. Marcus, J. B. Miller, 
and E. Rezayi. Many of the results discussed in this paper have been obtained independently by S.-S. Lee et 
al. \cite{LRNF2007} in a work which is being submitted simultaneously. This work was supported in part by 
NSF grant DMR-0541988, the Harvard Society of Fellows, and the Heisenberg program of DFG.

\newcommand{\noopsort}[1]{} \newcommand{\printfirst}[2]{#1}
  \newcommand{\singleletter}[1]{#1} \newcommand{\switchargs}[2]{#2#1}


\begin{thebibliography}{18}
\expandafter\ifx\csname natexlab\endcsname\relax\def\natexlab#1{#1}\fi
\expandafter\ifx\csname bibnamefont\endcsname\relax
  \def\bibnamefont#1{#1}\fi
\expandafter\ifx\csname bibfnamefont\endcsname\relax
  \def\bibfnamefont#1{#1}\fi
\expandafter\ifx\csname citenamefont\endcsname\relax
  \def\citenamefont#1{#1}\fi
\expandafter\ifx\csname url\endcsname\relax
  \def\url#1{\texttt{#1}}\fi
\expandafter\ifx\csname urlprefix\endcsname\relax\def\urlprefix{URL }\fi
\providecommand{\bibinfo}[2]{#2}
\providecommand{\eprint}[2][]{\url{#2}}

\bibitem[{\citenamefont{Moore and Read}(1991)}]{MR9162}
\bibinfo{author}{\bibfnamefont{G.}~\bibnamefont{Moore}} \bibnamefont{and}
  \bibinfo{author}{\bibfnamefont{N.}~\bibnamefont{Read}},
  \bibinfo{journal}{Nucl. Phys. B} \textbf{\bibinfo{volume}{360}},
  \bibinfo{pages}{362} (\bibinfo{year}{1991}).

\bibitem[{\citenamefont{Xia et~al.}(2004)\citenamefont{Xia, Pan, Vicente,
  Adams, Sullivan, Stormer, Tsui, Pfeiffer, Baldwin, and West}}]{XPV0409}
\bibinfo{author}{\bibfnamefont{J.}~\bibnamefont{Xia}},
  \bibinfo{author}{\bibfnamefont{W.}~\bibnamefont{Pan}},
  \bibinfo{author}{\bibfnamefont{C.~L.} \bibnamefont{Vicente}},
  \bibinfo{author}{\bibfnamefont{E.~D.} \bibnamefont{Adams}},
  \bibinfo{author}{\bibfnamefont{N.~S.} \bibnamefont{Sullivan}},
  \bibinfo{author}{\bibfnamefont{H.~L.} \bibnamefont{Stormer}},
  \bibinfo{author}{\bibfnamefont{D.~C.} \bibnamefont{Tsui}},
  \bibinfo{author}{\bibfnamefont{L.~N.} \bibnamefont{Pfeiffer}},
  \bibinfo{author}{\bibfnamefont{K.~W.} \bibnamefont{Baldwin}},
  \bibnamefont{and} \bibinfo{author}{\bibfnamefont{K.~W.} \bibnamefont{West}},
  \bibinfo{journal}{Phys. Rev. Lett.} \textbf{\bibinfo{volume}{93}},
  \bibinfo{pages}{176809} (\bibinfo{year}{2004}).

\bibitem[{\citenamefont{Greiter et~al.}(1991)\citenamefont{Greiter, Wen, and
  Wilczek}}]{GWW9105}
\bibinfo{author}{\bibfnamefont{M.}~\bibnamefont{Greiter}},
  \bibinfo{author}{\bibfnamefont{X.-G.} \bibnamefont{Wen}}, \bibnamefont{and}
  \bibinfo{author}{\bibfnamefont{F.}~\bibnamefont{Wilczek}},
  \bibinfo{journal}{Phys. Rev. Lett.} \textbf{\bibinfo{volume}{66}},
  \bibinfo{pages}{3205} (\bibinfo{year}{1991}).

\bibitem[{\citenamefont{Morf and d'Ambrumenil}(2003)}]{MN0309}
\bibinfo{author}{\bibfnamefont{R.}~\bibnamefont{Morf}} \bibnamefont{and}
  \bibinfo{author}{\bibfnamefont{N.}~\bibnamefont{d'Ambrumenil}},
  \bibinfo{journal}{Phys. Rev. B} \textbf{\bibinfo{volume}{68}},
  \bibinfo{pages}{113309} (\bibinfo{year}{2003}).

\bibitem[{\citenamefont{Rezayi and Haldane}(2000)}]{RH0085}
\bibinfo{author}{\bibfnamefont{E.~H.} \bibnamefont{Rezayi}} \bibnamefont{and}
  \bibinfo{author}{\bibfnamefont{F.~D.} \bibnamefont{Haldane}},
  \bibinfo{journal}{Phys. Rev. Lett.} \textbf{\bibinfo{volume}{84}},
  \bibinfo{pages}{4685} (\bibinfo{year}{2000}).

\bibitem[{\citenamefont{Kane and Fisher}(1997)}]{KF9732}
\bibinfo{author}{\bibfnamefont{C.~L.} \bibnamefont{Kane}} \bibnamefont{and}
  \bibinfo{author}{\bibfnamefont{M.~P.~A.} \bibnamefont{Fisher}},
  \bibinfo{journal}{Phys. Rev. B} \textbf{\bibinfo{volume}{55}},
  \bibinfo{pages}{15832} (\bibinfo{year}{1997}).

\bibitem[{\citenamefont{Kitaev}(2005)}]{K0538}
\bibinfo{author}{\bibfnamefont{A.}~\bibnamefont{Kitaev}},
  \bibinfo{journal}{cond-mat/0506438}  (\bibinfo{year}{2005}).

\bibitem[{\citenamefont{Wen}(1993)}]{W9355}
\bibinfo{author}{\bibfnamefont{X.-G.} \bibnamefont{Wen}},
  \bibinfo{journal}{Phys. Rev. Lett.} \textbf{\bibinfo{volume}{70}},
  \bibinfo{pages}{355} (\bibinfo{year}{1993}).

\bibitem[{\citenamefont{Kane et~al.}(1994)\citenamefont{Kane, Fisher, and
  Polchinski}}]{KFP9429}
\bibinfo{author}{\bibfnamefont{C.~L.} \bibnamefont{Kane}},
  \bibinfo{author}{\bibfnamefont{M.~P.~A.} \bibnamefont{Fisher}},
  \bibnamefont{and}
  \bibinfo{author}{\bibfnamefont{J.}~\bibnamefont{Polchinski}},
  \bibinfo{journal}{Phys. Rev. Lett.} \textbf{\bibinfo{volume}{72}},
  \bibinfo{pages}{4129} (\bibinfo{year}{1994}).

\bibitem[{\citenamefont{Giamarchi and Schultz}(1988)}]{GS8825}
\bibinfo{author}{\bibfnamefont{T.}~\bibnamefont{Giamarchi}} \bibnamefont{and}
  \bibinfo{author}{\bibfnamefont{H.~J.} \bibnamefont{Schultz}},
  \bibinfo{journal}{Phys. Rev. B} \textbf{\bibinfo{volume}{37}},
  \bibinfo{pages}{325} (\bibinfo{year}{1988}).

\bibitem[{\citenamefont{Wen}(1995)}]{W9505}
\bibinfo{author}{\bibfnamefont{X.-G.} \bibnamefont{Wen}},
  \bibinfo{journal}{Advances in Physics} \textbf{\bibinfo{volume}{44}},
  \bibinfo{pages}{405} (\bibinfo{year}{1995}).

\bibitem[{\citenamefont{Miller et~al.}(2007)\citenamefont{Miller, Radu,
  Zumbuhl, Levenson-Falk, Kastner, Marcus, Pfeiffer, and West}}]{MRZ2007}
\bibinfo{author}{\bibfnamefont{J.~B.} \bibnamefont{Miller}},
  \bibinfo{author}{\bibfnamefont{I.~P.} \bibnamefont{Radu}},
  \bibinfo{author}{\bibfnamefont{D.~M.} \bibnamefont{Zumbuhl}},
  \bibinfo{author}{\bibfnamefont{E.~M.} \bibnamefont{Levenson-Falk}},
  \bibinfo{author}{\bibfnamefont{M.~A.} \bibnamefont{Kastner}},
  \bibinfo{author}{\bibfnamefont{C.~M.} \bibnamefont{Marcus}},
  \bibinfo{author}{\bibfnamefont{L.~N.} \bibnamefont{Pfeiffer}},
  \bibnamefont{and} \bibinfo{author}{\bibfnamefont{K.~W.} \bibnamefont{West}},
  \bibinfo{journal}{Nature Physics (to appear)}  (\bibinfo{year}{2007}).

\bibitem[{\citenamefont{Lal}(2006)}]{L0618}
\bibinfo{author}{\bibfnamefont{S.}~\bibnamefont{Lal}},
  \bibinfo{journal}{cond-mat/0611218}  (\bibinfo{year}{2006}).

\bibitem[{\citenamefont{Nayak and Wilczek}(1996)}]{NW9629}
\bibinfo{author}{\bibfnamefont{C.}~\bibnamefont{Nayak}} \bibnamefont{and}
  \bibinfo{author}{\bibfnamefont{F.}~\bibnamefont{Wilczek}},
  \bibinfo{journal}{Nucl. Phys. B} \textbf{\bibinfo{volume}{479}},
  \bibinfo{pages}{529} (\bibinfo{year}{1996}).

\bibitem[{\citenamefont{Stern and Halperin}(2006)}]{SH0602}
\bibinfo{author}{\bibfnamefont{A.}~\bibnamefont{Stern}} \bibnamefont{and}
  \bibinfo{author}{\bibfnamefont{B.~I.} \bibnamefont{Halperin}},
  \bibinfo{journal}{Phys. Rev. Lett.} \textbf{\bibinfo{volume}{96}},
  \bibinfo{pages}{016802} (\bibinfo{year}{2006}).

\bibitem[{\citenamefont{Bonderson et~al.}(2006)\citenamefont{Bonderson, Kitaev,
  and Shtengel}}]{BKS0603}
\bibinfo{author}{\bibfnamefont{P.}~\bibnamefont{Bonderson}},
  \bibinfo{author}{\bibfnamefont{A.}~\bibnamefont{Kitaev}}, \bibnamefont{and}
  \bibinfo{author}{\bibfnamefont{K.}~\bibnamefont{Shtengel}},
  \bibinfo{journal}{Phys. Rev. Lett.} \textbf{\bibinfo{volume}{96}},
  \bibinfo{pages}{016803} (\bibinfo{year}{2006}).

\bibitem[{\citenamefont{Rezayi}(2007)}]{EHR2007}
\bibinfo{author}{\bibfnamefont{E.~H.} \bibnamefont{Rezayi}},
  \bibinfo{journal}{(private communication)}  (\bibinfo{year}{2007}).

\bibitem[{\citenamefont{Lee et~al.}(2007)\citenamefont{Lee, Ryu, Nayak, and
  Fisher}}]{LRNF2007}
\bibinfo{author}{\bibfnamefont{S.-S.} \bibnamefont{Lee}},
  \bibinfo{author}{\bibfnamefont{S.}~\bibnamefont{Ryu}},
  \bibinfo{author}{\bibfnamefont{C.}~\bibnamefont{Nayak}}, \bibnamefont{and}
  \bibinfo{author}{\bibfnamefont{M.~P.~A.} \bibnamefont{Fisher}},
  \bibinfo{journal}{cond-mat/0707.0478}  (\bibinfo{year}{2007}).

\end{thebibliography}
\end{document}